\documentclass[10pt,prc,aps]{revtex4}   
\usepackage{graphicx}
\tighten
\begin{document}
\draft
\title{ The quark-hadron phase transition in \\                                                     
 weakly isospin-asymmetric nuclear matter 
 }              
\author{            
 Francesca Sammarruca}      
\affiliation{                 
 Physics Department, University of Idaho, Moscow, ID 83844-0903, U.S.A  } 
\date{\today} 
\email{fsammarr@uidaho.edu}
\begin{abstract}
We consider the transition from quark to hadronic matter which may result during the cooling/expansion
of the quark-gluon plasma formed in energetic collisions of weakly asymmetric ions. 
This transition involves the energy density of $u$ and $d$ quark matter and the one of nearly 
isospin-symmetric nuclear matter. Within bag models, the former entails knowledge of the bag pressure, 
a poorly constrained quantity. The bag pressure at high-density can be fixed
imposing equality of quark and nucleonic energy densities at the (assumed known) transition point. 
We find this value to be very model dependent. 

\end{abstract}
\maketitle

\section{Introduction} 
                                                                     
The occurence of a transition between hadronic phase and (deconfined) quark phase in dense matter is an unsettled     
issue. From the experimental side, 
observing the formation of a quark-gluon plasma is one of the main objectives of CERN-SPS and RHIC experiments. This 
state of matter 
is also expected to take place in the interior of neutron stars, due to the high pressures typical of 
the stellar core. 
However, the phase transition in energetic heavy-ion experiments typically involves low densities and high temperatures,
whereas high density cold matter constitutes the interior of stable neutron stars. Thus, there is some uncertainty
as to whether laboratory observations of the quark-gluon plasma can be helpful in considerations of neutron star
structure. On the other hand, there are also indications \cite{Cley86} that the deconfined phase occurs at a nearly 
constant value of the energy density, regardless the thermodynamical status of the system. 
A value of approximately 1 GeV fm$^{-3}$ for the energy density has been reported from CERN-SPS experiments \cite{CERN},
which we will take as a reasonable estimate of the transition energy density.           

Quark matter is a Fermi gas of 3$A$ quarks which, together, constitute a single color-singlet baryon with 
baryon number $A$. Strange matter is quark matter where flavor equilibrium has been established by the 
weak interaction. 
Deconfined matter is typically handled within bag models, such as the MIT bag model \cite{MIT}. 
The energy density of quark matter contains the bag parameter $B$, which represents the difference between
the perturbative vacuum and the true vacuum. Physically, $B$ is a pressure which maintains the quarks at finite 
density and chemical potential \cite{MIT}. 

The purpose of this paper is to explore the characteristics and the model dependence of the quark-hadron phase transition in weakly isospin-asymmetric nuclear 
matter. 
The paper is organized as follows: In the next Section, we review the main steps in the calculation of 
the energy density of quark matter within the MIT bag model. In Section III, the phase transition is   
 approached as resulting from energetic heavy-ion collisions of $^{208}$Pb nuclei and subsequent expansion
of the plasma, namely, it is a 
quark to hadron transition, where the hadronic phase consists of 
nearly symmetric nuclear matter. 
Assuming (approximate) knowledge of the transition density, we discuss possible values of $B$ and 
a simple parametrization 
for its density dependence, based on the condition that the energy densities
of the quark phase and the hadronic phase are equal at the transition.                               
Often we will follow the procedure of Ref.~\cite{BBSS} closely, with the specific intent of comparing findings. In that study,
both the equation of state (EoS) based on the 
Brueckner-Hartree-Fock (BHF) approach implemented with three-body forces (TBF) \cite{BHF} and the one based on 
relativistic mean field models (RMF) \cite{RMF1,RMF2,RMF3} are considered. 
On the other hand, our                    
nucleonic equation of state is obtained within the Dirac-Brueckner-Hartree-Fock model (DBHF) \cite{FS10}. 
A recent discussion on differences/similarities (with respect to effective or explicit inclusion of TBF)
between the DBHF scheme and the BHF + TBF one can be found in Ref.~\cite{FS10}. 
It is the purpose of this note to detect differences (in the predicted features of the phase transition)
which may originate from the nucleonic EoS. This is an important point, in view of the large model
dependence still existing among predictions of the latter, in particular the details of its slope.         

Once the bag pressure is ``determined" (or, rather, constrained to be within some reasonable range
of values), one can go back to the formalism we outline in Section II and calculate, for instance, the energy density
of $\beta$-stable strange quark matter for the purpose of stellar structure calculations. 

Our conclusions are summarized in the last Section. 

\section{Composition of $\beta$-equilibrated strange matter: basic equations} 
We consider a degenerate Fermi gas of $u$, $d$, and $s$ quarks and electrons in chemical
equilibrium maintained by the weak processes:
\begin{equation}
u + e^- \rightarrow d+ \nu_e  \; ,  
\end{equation}
\begin{equation}
u + e^- \rightarrow s+ \nu_e  \; ,  
\end{equation}
\begin{equation}
d \rightarrow u+ e^- + {\bar \nu_e}  \; ,  
\end{equation}
\begin{equation}
s \rightarrow u+ e^- + {\bar \nu_e}  \; ,  
\end{equation}
\begin{equation}
s +u \rightarrow d + u \; .                            
\end{equation}
The neutrinos are ignored as they are expected to have no impact on the dynamics, being an extremely diluted
gas (although massive neutrinos could be found in strange matter \cite{MIT}). 
In chemical equilibrium, we have 
\begin{equation} 
\mu _d = \mu_s = \mu          \; ,  
\end{equation}
\begin{equation} 
\mu _u +\mu_e = \mu                  \; .  
\end{equation}
Also, charge neutrality and baryon number conservation require
\begin{equation} 
\frac{2}{3}\rho_u=\frac{1}{3}\rho_d+\frac{1}{3}\rho_s+\rho_e \; , 
\end{equation}
and 
\begin{equation} 
\rho=\frac{1}{3}(\rho_d+\rho_u+\rho_s) \; , 
\end{equation}
where $\rho$ is the total (fixed) baryon density. 
Exploiting the relation between the density of each species and the corresponding thermodynamic potential,
\begin{equation} 
 \rho_i = -\frac{\partial \Omega_i }{\partial \mu_i }  \; , \; \; \; \; \; \; (i=u,d,s,e)
\end{equation}
equations~(6-9) can be solved for the chemical potentials of each species (and thus their densities
or fractions). 

To first order in $a_c$ (the QCD coupling constant), the thermodynamic potentials 
are \cite{MIT} 
\begin{equation} 
 \Omega_u = -\frac{\mu_u ^4}{4 \pi^2} \Big(1-\frac{2 a_c}{\pi}\Big)            \; , 
\end{equation}
\begin{equation} 
 \Omega_d = -\frac{\mu_d ^4}{4 \pi^2}\Big (1-\frac{2 a_c}{\pi}\Big)            \; , 
\end{equation}

\begin{widetext}
\begin{eqnarray}
\Omega_s &=&
-\frac{1}{4\pi^2}\Big(\mu_s(\mu_s^2 - m_s^2)^{1/2} (\mu_s^2 -\frac{5}{2}m_s^2)
+ \frac{3}{2}m_s^4 ln\frac{\mu_s + (\mu_s^2 - m_s^2)^{1/2}}{m_s} \nonumber\\
&-&\frac{2 a_c}{\pi}\Big (3 \Big ( \mu_s (\mu_s^2 - m_s^2)^{1/2}-m_s^2 
ln\frac{\mu_s + (\mu_s^2 - m_s^2)^{1/2}}{m_s}\Big)^2 -2(\mu_s^2 -m_s^2)^{1/2}+3 m_s^4ln^2\frac{m_s}{\mu_s}
\nonumber\\
&+& 6 ln\frac{\rho}{\mu_s} \Big (\mu_s m_s^2(\mu_s^2 - m_s^2)^{1/2} -m_s^4
ln\frac{\mu_s + (\mu_s^2 - m_s^2)^{1/2}}{m_s}                                                                
\Big) \Big ) \Big ) \; . 
\end{eqnarray}
\end{widetext}

The total energy density is then
\begin{equation} 
\epsilon = \sum _{i=u,d,s,e} (\Omega_i + \mu_i\rho_i) + B                                \; , 
\end{equation}
where $B$ is a positive energy per unit volume carried by the vacuum, that is, the vacuum pressure. 
Its value is an open question, and will be confronted in the next section. 

Quark masses are taken to be equal to their current-algebra values, which amounts to ignoring effects
from chiral symmetry breaking (i.~e. dynamical masses) in the quark gas. In particular, $u$ and $d$
quarks can be assumed to be massless, whereas a mass between 100 and 200 MeV is typically assigned             
to the $s$ quark.

\section{The quark-hadron phase transition in nearly isospin-symmetric matter } 
If lead nuclei are accelerated at CERN-SPS energies, during the early stages of the collision 
the so-called ``fireball" is formed, which is a hot and dense plasma of deconfined quarks and gluons.
As the plasma cools down, it becomes more diluted until hadronization takes place. Thus, in the 
phase diagram one is moving from higher to lower densities, and from the quark to the hadronic phase. 
Weak processes are not expected to play a role due to the short expansion times, and thus strangeness, which
is conserved in strong interactions, can be ignored.
 The bulk of nuclear matter resulting after hadronization 
can be described by the EoS of cold matter with an average isospin asymmetry, $\alpha$, equal to 0.2
($\alpha$ is defined as $(N-Z)/(N+Z)$, and 0.2 is its value in lead)). For the hadronic phase, we then have, for the energy density,
\begin{equation} 
\epsilon_H = (e (\rho, \alpha) + x_p m_p + x_n m_n)\rho \; ,                                        
\end{equation}
where $e$ is the energy per particle from Ref.~\cite{FS10} and $x_{p}$ ($x_{n}$) are the proton (neutron) fractions. 

Taking the transition energy density, $\epsilon^{tr}$, to be known, we can immediately determine the baryonic density,
$\rho^{tr}$, which corresponds 
to it in our model of the nucleonic EoS. For instance, for $\epsilon^{tr} \approx 0.8 GeV fm^{-3}$,
the transition baryon number density is 0.74 fm$^{-3}$ from our EoS.                   
Notice that 
the stiffness of the nucleonic EoS determines the {\it baryonic} density at which the         
appropriate transition energy density is achieved, a point worth underlining.  

The bag pressure is essentially a free parameter. It has a value of 55 MeV fm$^{-3}$ according to the original
MIT bag model, whereas lattice calculations estimate it to be about 
200 MeV fm$^{-3}$ \cite{lattice}.   
First, we consider (constant) values of $B$ within this (rather large) range and calculate the energy density
of $u$ and $d$ quark matter. In Fig.~1, we show the quark energy density for $B$ equal to 
55, 100, and 200 MeV fm$^{-3}$, and the nucleonic energy density from Eq.~(15). 
The two panels refer to $a_c$ equal to 0 and 0.2, respectively. Referring to the first
panel, we see that, for $B$ 
equal to 55 MeV fm$^{-3}$, the crossing density is too low to be realistic (about normal nuclear
matter density), 
whereas with values of 100 and 200 MeV fm$^{-3}$ transition densities equal to 0.93 and 1.2 fm$^{-3}$ are               
predicted, respectively. The corresponding energy densities are not inconsistent with present
experimental constraints. When $a_c$ is non-zero, the transition density shifts to higher values,
see second frame of Fig.~1, but the difference is minor. 

\begin{figure}[!t] 
\centering 
\vspace*{1.0cm}
\hspace*{-0.5cm}
\scalebox{0.39}{\includegraphics{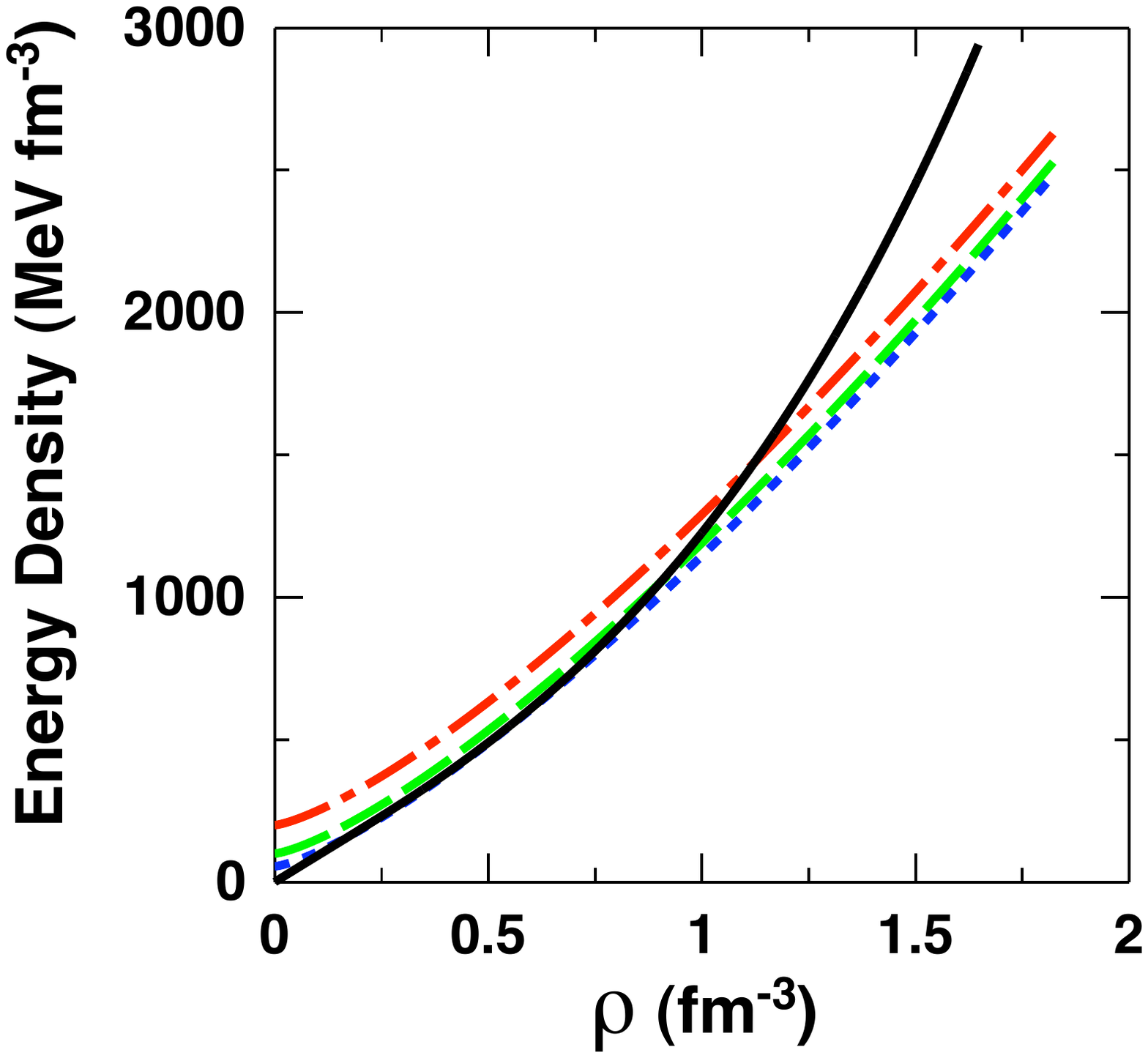}} 
\vspace*{3.0cm}
\hspace*{1.2cm}
\vspace*{0.5cm}
\hspace*{-1.0cm}
\scalebox{0.39}{\includegraphics{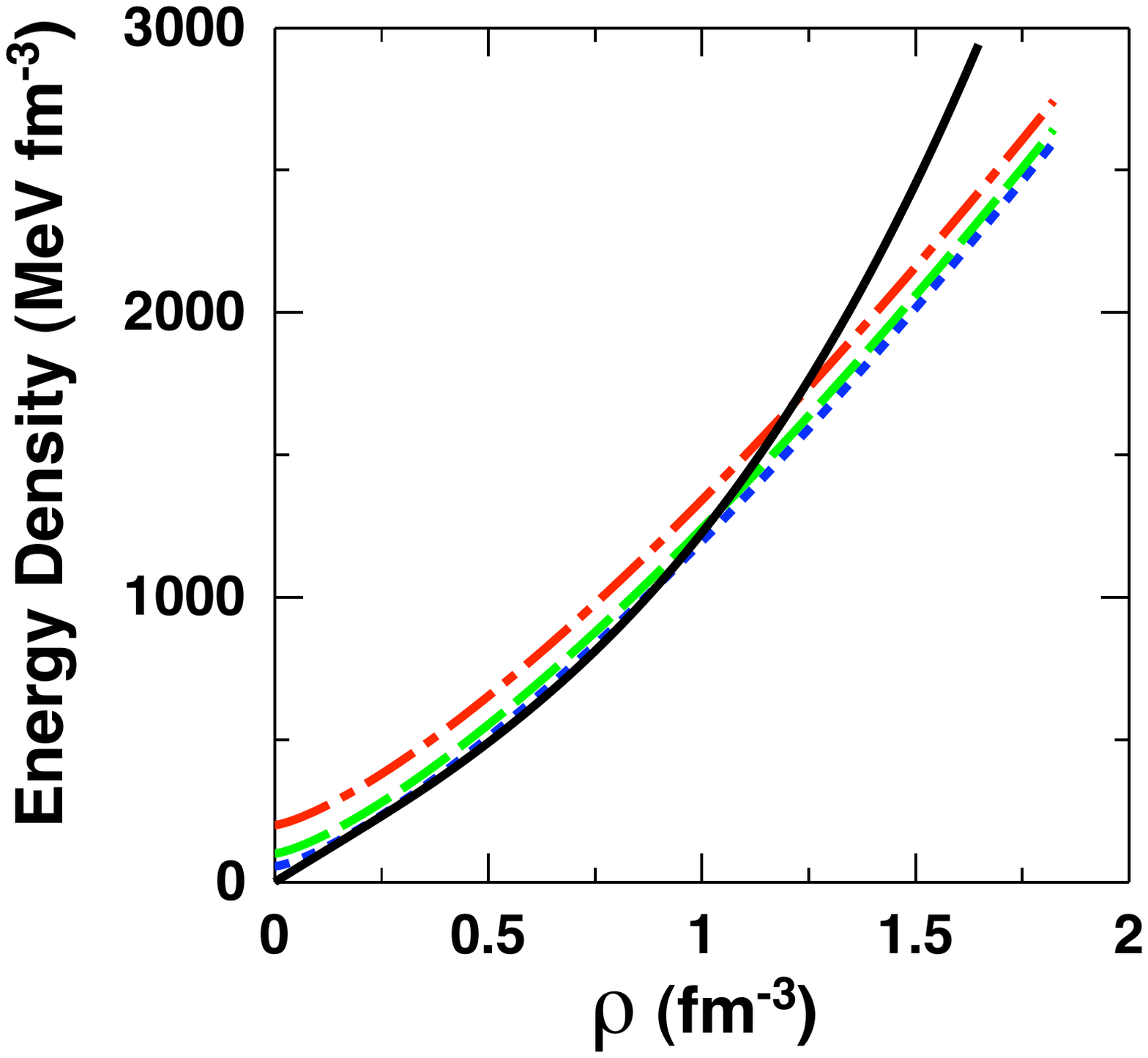}} 
\vspace*{-5.5cm}
\caption{(color online) The hadronic (solid line) and the quark energy densities for different values 
of the bag constant. Dotted line: $B$= 55 MeV $fm^{-3}$; 
Dashed line: $B$= 100 MeV $fm^{-3}$; 
Dash-dotted line: $B$= 200 MeV $fm^{-3}$. 
The left and right frames refer to values of the QCD coupling constant equal to 0 and 0.2,        
respectively. 
}
\label{one}
\end{figure}

A better way to proceed is to fit the value of $B$ imposing equality of the quark and hadron energy
densities at the appropriate baryon number density (as determined from the nucleonic EoS).
That is: 
\begin{equation}
\epsilon_H^{tr} = \epsilon_q^{tr} \; , 
\end{equation}
where the right-hand side 
is the energy density of $u$ and $d$ quarks, which depend linearly on $B$, see Eq.~(14). With the left-hand side known
from the nucleonic EoS, it is elementary to determine $B$. 
It has been argued, though, that $B$ should be density dependent \cite{JJ}, as a large reduction of its value in nuclear
matter appears to be consistent with the description of phenomena such as the EMC effect \cite{JJ}. 
In fact, allowing the MIT bag constant to depend on the local density leads to the recovery of relativistic
phenomenology \cite{JJ} like large cancelations between Lorentz scalar and vector potentials in nuclear matter.

Adopting an even simpler parametrization than those considered in Ref.~\cite{BBSS}, we 
assume that $B(\rho)$ has a constant value up to densities around the transition point and 
then drops sharply to its asymptotic value 
(basically, a very sharp Woods-Saxon distribution approximated by a step function). 
That is: 
\begin{equation}
B(\rho)=\left\{
\begin{array}{l l}
B_0 & \quad \mbox{if $\rho < \rho^{tr}\; ,$ }\\
B_{asy} & \quad \mbox{otherwise.}
\end{array}
\right.
\end{equation}
Obviously, 
high-density predictions will be mostly sensitive to the asymptotic value of $B(\rho)$, thus the 
value of $B_0$ is not very relevant. We follow Ref.~\cite{BBSS} and choose a value between 200 and 400 MeV fm$^{-3}$.
 $B_{asy}$ can then be fixed by imposing the condition Eq.~(16).                                 
For $a_c$=0, fitted values of $B_{asy}$ are shown in Table I for some acceptable values of the 
transition energy density. 
Our values of $B_{asy}$  
are directly comparable with those of the constant $B_{\infty}$ of Ref.~\cite{BBSS}, where a Gaussian or Woods-Saxon
parametrizations are chosen for $B(\rho)$, 
with its asymptotic value, $B_{\infty}$, fitted at the transition point. That is the parameter displayed
in Table I for the second and third models. 

First, we notice that, within a given model, 
$B_{asy}$ spans a rather large range (for relatively small changes in the transition density). 
Furthermore, the values of $B$ are quite sensitive to the details 
of the density dependence of the nucleonic EoS. Clearly, some (pre-established) value of the energy density
is obtained at a lower or higher value of the number density depending on the stiffness of the original EoS.
This becomes particularly clear looking at the last model considered in Table I, which is 
 based on the ``BHF + TBF" approach, but is considerably more repulsive. This EoS is the one labeled as ``BOB" in 
Ref.\cite{BOB}. It is obtained with BHF calculations together with TBF. The parameters of the TBF are 
chosen to be, as far as possible, consistent with those of the meson-exchange two-body potential, which is 
Bonn B \cite{Mac89}.

\begin{table}                
\centering \caption                                                    
{Values of $B_{asy}$ (or $B_{\infty}$) predicted by various hadronic models for fixed transition energy   
density. The ``BHF + TBF" and ``RMF" entries are taken from Ref.~\cite{BBSS}. (For those cases, the given values 
refer to the parameter labeled as $B_{\infty}$ in Ref.~\cite{BBSS}.)
The ``BHF + TBF$_2$" is the EoS model labeled as ``BOB" in Ref.~\cite{BOB}, see text for. 
} 
\vspace{5mm}

\begin{tabular}{|c|c|c|c|}
\hline

Model  & $\epsilon_{tr}$ (MeV fm$^{-3}$) & $\rho_{tr}$ (fm$^{-3}$) &
                    $B_{asy}$ (MeV fm$^{-3}$) \\                                        
\hline
DBHF &   0.8 $\times 10^{3}$  & 0.74  &  35.7                \\                  
     &   1.1 $\times 10^{3}$  & 0.93  &  58.9                \\                  
BHF+TBF     &   0.8 $\times 10^{3}$  & 0.76  &  36.4                \\                  
            &   1.1 $\times 10^{3}$  & 0.97  &  51.1                \\                  
RMF         &   0.8 $\times 10^{3}$  & 0.76  &  37.9                \\                  
            &   1.1 $\times 10^{3}$  & 0.98  &  37.8                \\                  
(BHF+TBF)$_2$     &   0.8 $\times 10^{3}$  & 0.73  &  49.4                \\                  
            &   1.1 $\times 10^{3}$  & 0.89  &  122.4               \\                  
\hline

\end{tabular}
\end{table}

\begin{figure}[!t] 
\centering 
\scalebox{0.41}{\includegraphics{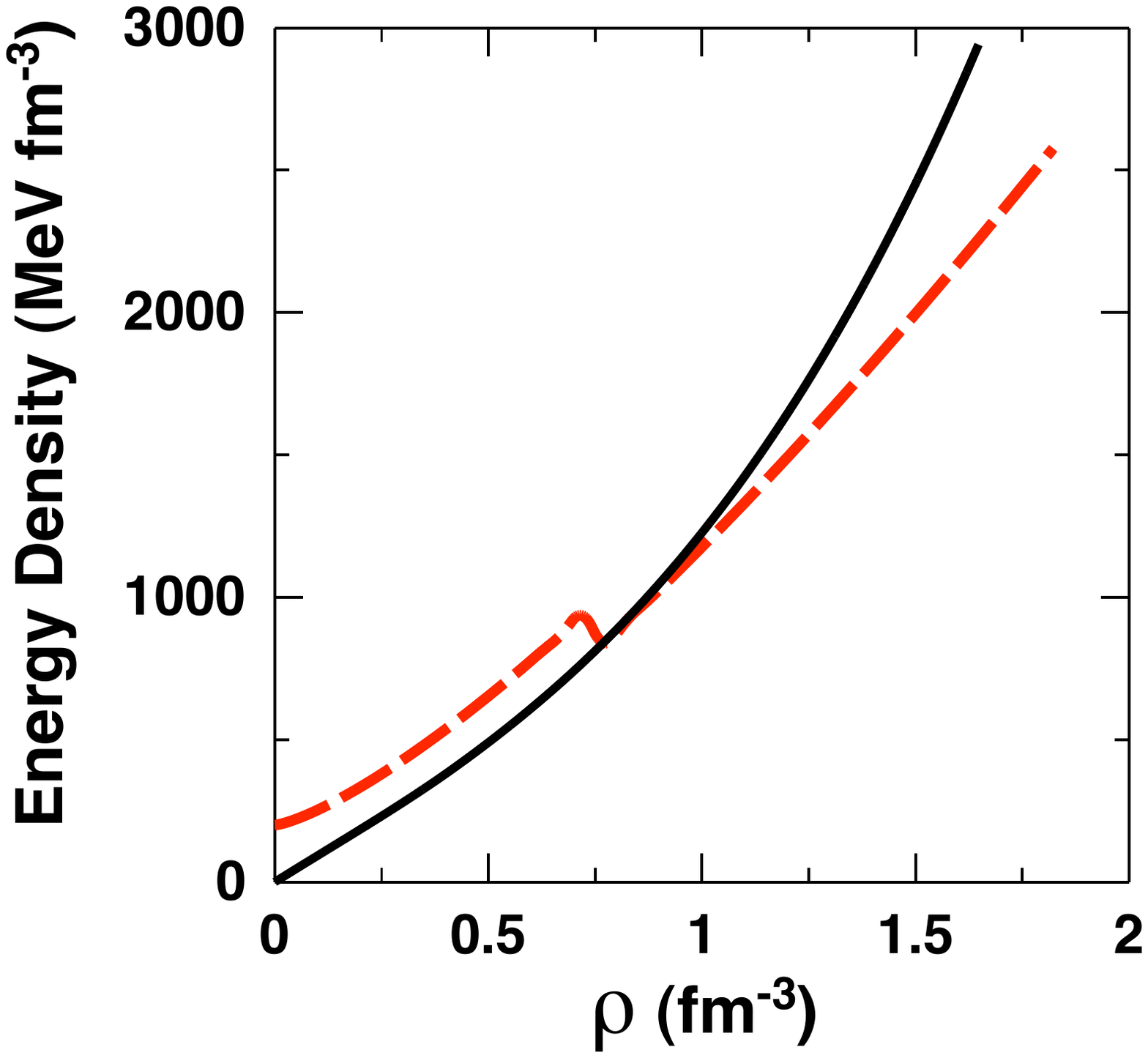}} 
\hspace*{-1.0cm}
\scalebox{0.41}{\includegraphics{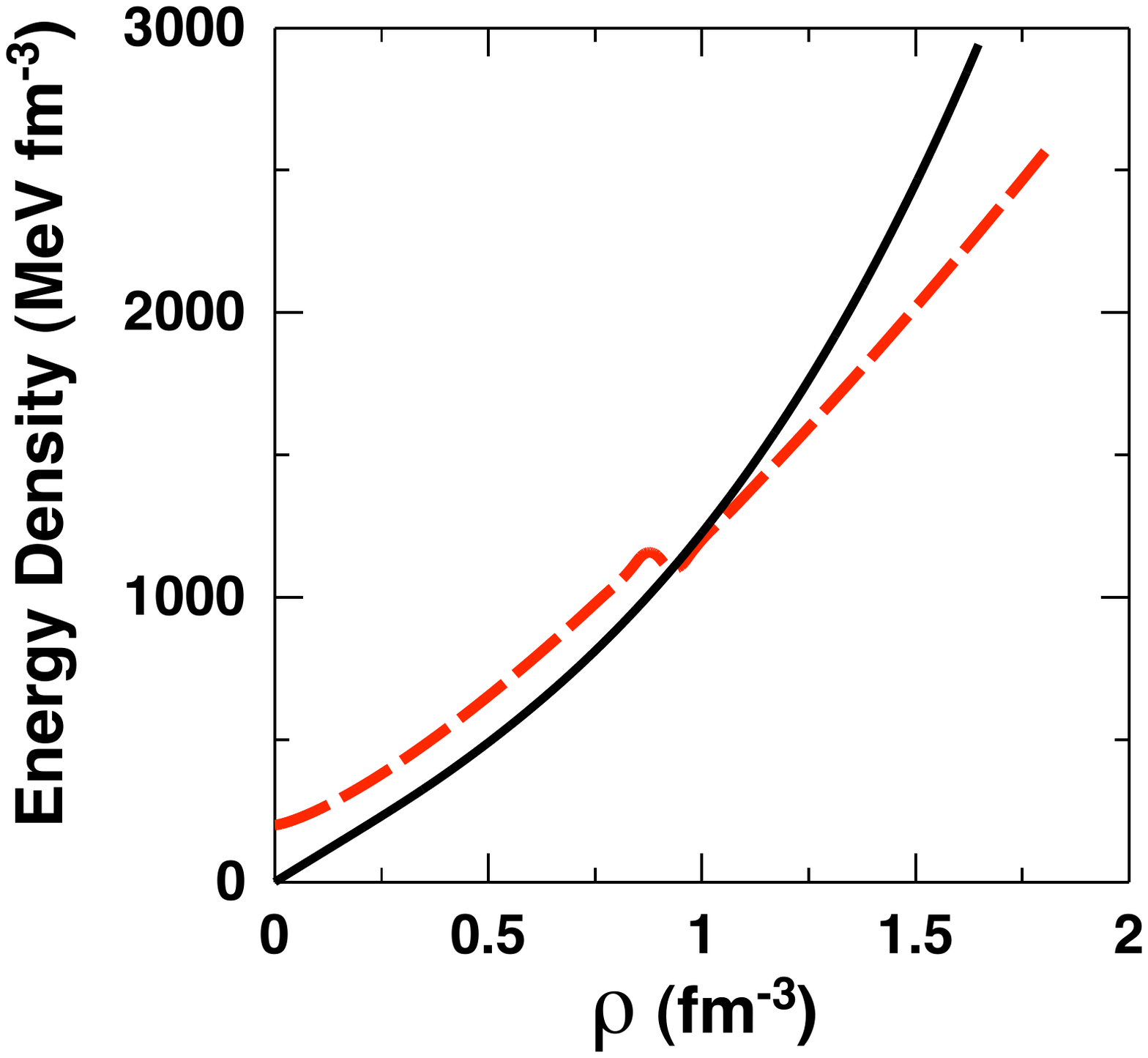}} 
\vspace*{-3.0cm}
\caption{(color online) The hadronic (solid line) and the quark (dashed line) energy densities with a density dependence
of the bag constant as in Eq.~(17). In the left (right) panel a transition energy density of 
800 (1100) MeV $fm^{-3}$ is assumed. 
}
\label{two}
\end{figure}

The quark and hadron energy densities with the sharp density dependence of $B(\rho)$ from Eq.~(17) are shown in Fig.~2, 
for transition energy density equal to 800 and 1100 MeV $fm^{-3}$. 
A smoother density dependence of $B(\rho)$ would have impact only at the lower densities.
Constructing one is straightforward but besides the point of this paper, where we want to perform a 
pedagogical demonstration of the uncertainty involved in determining the features of the phase transition 
(and the EoS thereafter), 
particularly as it relates to the underlying hadronic model. 

\section{Conclusions} 
We explored the quark-hadron phase transition in weakly isospin-asymmetric matter as it could take place during 
the expansion phase of energetic collisions of lead nuclei. 
We used either a constant 
bag pressure, $B$, or the simplest 
parametrization of its density dependence to demonstrate the role of the 
nucleonic EoS in determining the features of the transition. 
The parameter which determines the asymptotic behavior of $B(\rho)$ is very sensitive to the details
of the density dependence of the nucleonic EoS. Thus, independent information on the density dependence 
of $B$ in the medium would be very helpful. Those could come, for instance, from considerations of EMC 
effects \cite{JJ}. In turn, such information would help constrain the density dependence of the {\it nucleonic}
EoS at the phase transition.

\section*{Acknowledgments}
Support from the U.S. Department of Energy under Grant No. DE-FG02-03ER41270 is 
acknowledged.

\end{document}